\documentclass[prl,twocolumn,showpacs,amsmath,amssymb]{revtex4-1}

\usepackage{graphics,graphicx}

\begin{document}
\title{Strong nonlinear optical response of graphene flakes \\
measured by four-wave mixing}
\author{E.~Hendry$^{1\ast}$}
\author{P.~J.~Hale$^1$}
\author{J.~J.~Moger$^1$}
\author{A.~K.~Savchenko$^1$}
\author{S. A. Mikhailov$^2$}
\affiliation{1 School of Physics, University of Exeter, EX4 4QL, U.K.\\2 Institute of Physics, University of Augsburg, D-86135 Augsburg, Germany}

\pacs{}

\begin{abstract}
We present the first experimental investigation of nonlinear optical properties of graphene flakes. We find that at near infrared frequencies a graphene monolayer exhibits a remarkably high third-order optical nonlinearity which is practically independent of the wavelengths of incident light. The nonlinear optical response can be utilized for imaging purposes, with image contrasts of graphene which are orders of magnitude higher than those obtained using linear microscopy.
\end{abstract}

\maketitle

Graphene, a single sheet of carbon atoms in a hexagonal lattice, is the basic building block for all graphitic materials. Although it has been known as a theoretical concept for some time \cite{1}, a layer of graphene has only recently been isolated from bulk graphite and deposited on a dielectric substrate \cite{2}. The great interest in studying graphene is driven by its unique band structure and many unusual electrical, thermal, and mechanical properties \cite{3,5}. An important stage in the development of graphene-based structures was the demonstration that a monolayer of atoms can be visible under an optical microscope \cite{6,7,8,8a}. There have also been demonstrations of unusual optical properties of graphene: for example, a wavelength independent absorption of 2.3\% per graphene layer \cite{9,9a,9b}. Recently, it was predicted that graphene should demonstrate strongly nonlinear optical behavior at THz frequencies \cite{10}, which could lead to important device applications.

Here we present the first experimental studies of nonlinear optical properties of graphene at visible and near infrared frequencies. It is shown that graphene produces strong third-order nonlinear optical response, described by nonlinear susceptibility $|\chi^{(3)}|\sim10^{-7}$ esu (electrostatic units), which is comparable to that of other strongly nonlinear materials, such as carbon nanotubes \cite{12,13,14,15,16}. In contrast to carbon nanotubes \cite{16}, however, this nonlinear response is essentially dispersionless over the wavelength range in our experiments (emission with wavelength of 760 - 840 nm). We also show that the nonlinear response of graphene can be exploited for high-contrast optical imaging of graphene flakes: the image contrasts for monolayer flakes on a dielectric substrate are several orders of magnitude higher than those in reflection microscopy \cite{6,7,8}.

\begin{figure}[htb]{}
\includegraphics[width=.7\columnwidth]{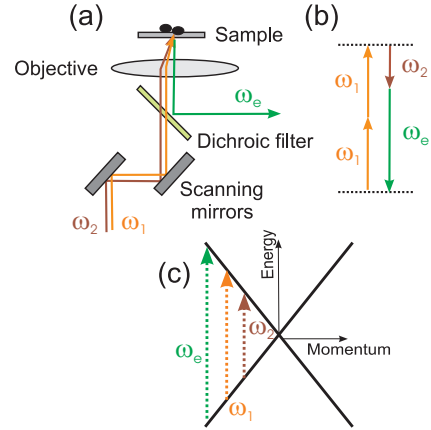}
\caption{(a) Schematic of the experimental layout, indicating the pump beams with frequencies $\omega_{1}$, $\omega_{2}$, and the emission beam with frequency $\omega_{e}$. (b) Diagram of energy conservation in four-wave process. (c) Band structure of graphene with the three photon energies (arrows) involved in four-wave mixing. }\label{fig1}
\end{figure}

To investigate the nonlinear response of graphene, we employ the four-wave mixing technique \cite{17}. Figure \ref{fig1}(a) illustrates the principle of the method: two incident pump laser beams with wavelengths $\lambda_{1}$ (tunable from 670 nm to 980 nm) and  $\lambda_{2}$ (1130 nm to 1450 nm) are focused collinearly onto a sample and mix together to generate a third, coherent beam of wavelength  $\lambda_{e}$. In our experiment, the pump beams are generated by an Optical Parametric Oscillator which results in collinear 6 ps pulses which overlap in time. The incident pump pulses are focused onto the sample using a water immersion objective with a numerical aperture of 1.2, giving rise to excitation powers at the sample of $\sim1$ mW.

\begin{figure}[htb]{}
\includegraphics[width=.7\columnwidth]{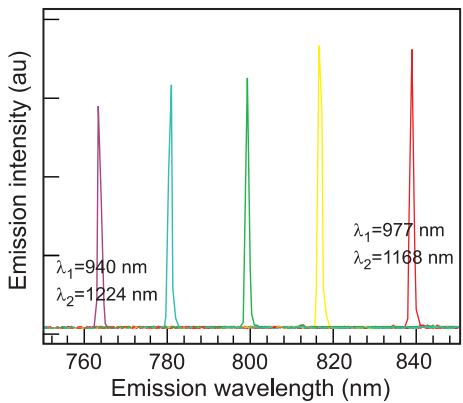}
\caption{Emission spectra of a graphene flake excited with pump light of different wavelengths, ($\lambda_{1}$, $\lambda_{2}$): (940 nm, 1224 nm), (950 nm, 1210 nm), (958 nm, 1196 nm), (967 nm, 1183 nm) and (977 nm, 1168 nm). }\label{fig2}
\end{figure}

Imaging is achieved by raster scanning of the excitation beams over the sample and acquiring the emitted signal as a function of its position. Images consisting of 512 by 512 pixels were acquired with a pixel dwell time of 2.6 $\mu$s resulting in an image frame rate of approximately 0.6 seconds. The nonlinear emission is collected in the backward direction using a 750 nm long-pass dichroic mirror followed by a 750 nm band pass filter (210 nm bandwidth), used to isolate the nonlinear emission from the pump beams. The emission is detected by either a red-sensitive photomultiplier tube (for imaging) or by a spectrometer (see Ref. \cite{18} for more details of the experimental setup).

Centro-symmetric materials, such as isolated sheets of graphene, do not possess second-order optical nonlinearities. The third-order nonlinearity of a material is described by the polarization
\begin{eqnarray}
P^{(3)}(\omega_{e})=\chi^{(3)}E(\omega_{1})E(\omega_{2})E(\omega_{3})\label{eqn:one},
\end{eqnarray}
where $\omega_{1}$, $\omega_{2}$ and $\omega_{3}$ are the frequencies of the electric field, $E$, of the incident beams. In the degenerate four-wave mixing process used in our experiments the two mixing frequencies are equal: $\omega_{1}=\omega_{3}$. Due to energy conservation, the frequency of emitted light is $\omega_{e}=2\omega_{1}-\omega_{2}$, Figure \ref{fig1}(b). The degree of optical nonlinearity of a sample is characterized by its third-order susceptibility, $\chi^{(3)}$. Dielectric materials (e.g. glasses) exhibit very low optical nonlinearities, with $|\chi^{(3)}|<10^{-14}$ esu \cite{19}. Conducting materials can have significantly higher nonlinear optical responses: for example, free electrons in narrow-gap III-V semiconductors give rise to some of the highest nonlinearities, with $|\chi^{(3)}|\sim10^{-2}$ esu \cite{20}. It is also known that large optical nonlinearities can arise when the mixing or emission frequencies ($\omega_{1}$, $\omega_{2}$ or $\omega_{e}$) match the energy difference between electronic levels \cite{16}. It is interesting to study the magnitude and frequency dependence of the nonlinear response of graphene, a gapless semiconductor with linear dispersion relation, Figure \ref{fig1}(c).

Graphene and few-layer graphene samples are fabricated using the standard method of mechanical exfoliation \cite{2} and deposited onto a 100 $\mu$m thick glass cover slip. Prior to investigation in the nonlinear microscope, the layer thickness is estimated via contrast measurements under an optical microscope using a method similar to Ref. \cite{8}. (We calculated the expected reflection coefficient of a system with $N$ layers assuming the dielectric constant of each layer to be that of graphite.) Figure \ref{fig2} shows the measured nonlinear signal from a monolayer flake as a function of emission wavelength $\lambda_{e}$ for several combinations of pump wavelengths $\lambda_{1}$ and $\lambda_{2}$. In all measurements we observe a clear spike in emission at the wavelength corresponding to the condition $\omega_{e}=2\omega_{1}-\omega_{2}$. When the intensity of the pump pulses is varied, the amplitude of the emission peak shows a cubic dependence on the intensity, which confirms the third-order nature of the response.

By changing the pump wavelengths, one can investigate the dispersion of the nonlinear signal. We have found that, for equivalent pump intensities, the amplitude of the emission only changes by $\sim 10\%$ over the range of incident pump wavelengths. The wavelength range that we use is limited at short wavelengths by the restriction that $\omega_{1}-\omega_{2}$ must be smaller than the frequency of the 2D Raman peak in graphene in order to avoid coherent anti-Stokes Raman scattering \cite{21}. On the long wavelength side the range is limited by the filters used in the experimental set up. We believe that the wavelength independence of the nonlinear signal is the result of the dispersion relation of graphene, where all photon energies in the four-wave mixing process can match electronic transitions, Figure \ref{fig1}(c).

\begin{figure}[htb]{}
\includegraphics[width=.8\columnwidth]{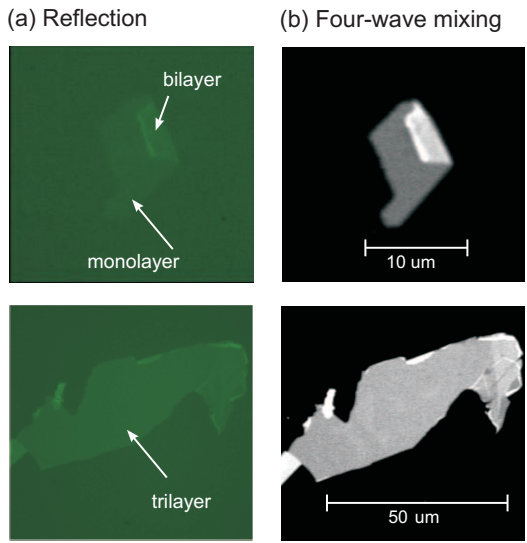}
\caption{(a) Standard green light reflection images of two graphene flakes. (b) Nonlinear optical images measured with pump wavelengths of 969 nm and 1179 nm.}\label{fig3}
\end{figure}

In Figure \ref{fig3} we compare the images obtained by measuring the four-wave mixing intensity to those obtained on the same flakes in an optical reflection microscope using 550 nm light (the standard optical imaging technique for thin graphene flakes \cite{7,8}). One can see a striking difference in the visibility of the flakes, which are significantly enhanced in the nonlinear image. This difference can be quantified in terms of the image contrast, defined as
\begin{eqnarray}
C=\frac{I_{gr}-I_{sub}}{I_{sub}}\label{eqn:two},
\end{eqnarray}
where $\emph{I}_{sub}$ and $\emph{I}_{gr}$ represent the signal arising from the substrate and the graphene flake, respectively. As the dielectric substrate does not show any measurable nonlinear signal, this contrast is very large and limited only by the noise on the detector: for a monolayer $C = 1.7\times10^{6}$ compared to $C = 0.08$ for standard optical reflection microscopy \cite{8}. (Similar enhanced imaging of carbon nanotubes using nonlinear optics has been recently observed in Ref. \cite{22}). We expect similarly high contrast for graphene on most dielectric substrates, since these materials have very weak optical nonlinearity \cite{19}. This makes the method of visualizing graphene flakes using nonlinear optical response rather universal.

\begin{figure}[htb]{}
\includegraphics[width=.8\columnwidth]{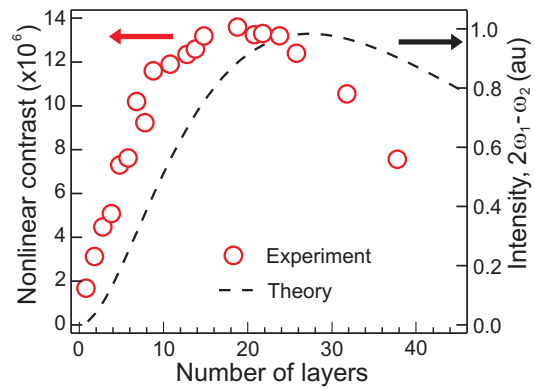}
\caption{The contrast in four-wave mixing images as a function of the number of graphene layers.}\label{fig4}
\end{figure}

Using different flakes, we have investigated the behavior of the nonlinear contrast as the number of graphene layers is increased. In Figure \ref{fig4} we plot the observed nonlinear image contrast as a function of the number of graphene layers, which is established from reflection microscopy \cite{8}. The nonlinear contrast is very sensitive to the number of graphene layers, $N$. It linearly increases with increasing the thickness up to $N\sim20$, while with further increase of $N$ the signal decreases towards zero, making thick ($> 70 \mathrm{\AA}$) flakes appear dark in the nonlinear images.


We have modelled this thickness dependence by considering a set of $N$ parallel graphene layers, each characterized by the linear-response conductivity $\sigma(\omega)$ and the third-order nonlinearity $\chi^{(3)}$. Solving the problem of the transmission of light through such a system we get the standard expression for the transmission amplitude $t(\omega)=[1+N2\pi\sigma(\omega)/c]^{-1}$ and the electric field at each graphene layer $E=E_0t(\omega)$, were $E_0$ is the amplitude of the incident wave. (We assume that the total thickness of the $N$-layered system is smaller than the wavelength of light.) The nonlinear current $j^{(3)}_{2\omega_1-\omega_2}$ induced in each layer by the electric field harmonics $\omega_1$ and $\omega_2$ is then proportional to
$\chi^{(3)}E_{\omega_1}^2E_{\omega_2} t^2(\omega_1) t(\omega_2)$.
This current generates an electromagnetic wave at the frequency $2\omega_1-\omega_2$ with the electric field amplitude $E_{2\omega_1-\omega_2}\propto j^{(3)}_{2\omega_1-\omega_2} Nt(2\omega_1-\omega_2)$ and the intensity
\begin{eqnarray}
I_{ 2\omega_1-\omega_2}\propto
|\chi^{(3)}|^2 I_{\omega_1}^2I_{\omega_2}N^2 | t^2(\omega_1)t(\omega_2)t(2\omega_1-\omega_2)|^2. \label{eqn:3}
\end{eqnarray}
Equation (\ref{eqn:3}) describes the nonlinear response of the system at the frequency $2\omega_1-\omega_2$. If one assumes that the optical conductivity of graphene is dispersionless and universal, $\sigma(\omega)\approx e^2/4\hbar$ \cite{9,9a,9b}, equation (\ref{eqn:3}) gives $I_{ 2\omega_1-\omega_2}\propto N^2/(1+N\pi e^2/2\hbar c)^8$. This model reproduces the trend in the nonlinear thickness dependence on $N$ with no fit parameters, Figure \ref{fig4}. The saturation and subsequent decrease of the nonlinear signal as a function of number of layers arises from the decrease of electric field inside the material for thicker samples. The difference in the exact positions of the maxima in experimental and theoretical curves could potentially be accounted for by taking into account the finite imaginary part of the conductivity.

\begin{figure}[htb]{}
\includegraphics[width=.8\columnwidth]{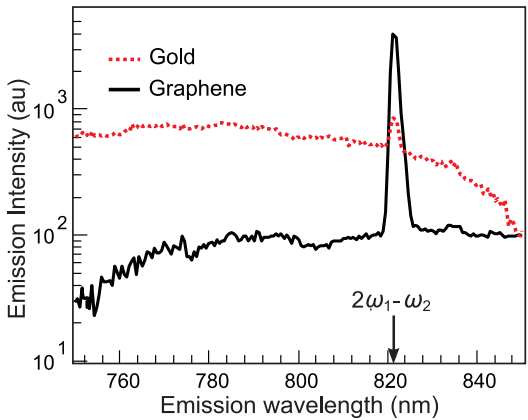}
\caption{Emission spectra of a flake excited with pump wavelengths (969 nm, 1179 nm), compared to the emission of a 4 nm thick gold film under the same experimental conditions.}\label{fig5}
\end{figure}

To quantify the magnitude of the observed nonlinear response in graphene, we compared the signal from monolayer graphene with that of a well characterized, quasi 2D material: a thin film of gold evaporated onto a glass microscope cover slip. The experiment with the gold film was performed in the same experimental conditions. It is known that a strong nonlinear response of thin gold films arises from the excitation of localized plasmons \cite{24,25}. In Figure \ref{fig5}, we present the comparison between the nonlinear emission of a 4 nm thick gold film with that from a monolayer graphene flake. One can see that the nonlinear emission from the gold film is very different from that in graphene. For graphene we observe a large peak at the emission wavelength, while in gold one can see a significantly smaller peak on top of a smooth background caused by two-photon luminescence in gold \cite{26}. To compare the magnitudes of third-order susceptibilities, one has to compare the magnitudes of the peaks, $\emph{I}_{gr}$ and $\emph{I}_{Au}$, with respect to the background. For the results presented in Figure \ref{fig5}, the ratio of the peak intensities in the two materials is approximately ten. From this, the nonlinear susceptibility of graphene, $\chi^{(3)}_{gr}$, can be found from that of the gold film ($|\chi^{(3)}_{Au}|\sim10^{-9}-10^{-8}$ esu \cite{24,25}) using the relation \cite{15,24}
\begin{eqnarray}
\frac{|\chi^{(3)}_{gr}|}{|\chi^{(3)}_{Au}|}\approx\frac{L_{gr}}{L_{Au}}\sqrt\frac{I_{gr}}{I_{Au}}\label{eqn:three},
\end{eqnarray}
where $\emph{L}_{gr}$ and $\emph{L}_{Au}$ is the thickness of the graphene flake and the gold film, respectively, and \emph{I} is the measured intensity of the four-wave signal. Given that $L_{Au}\sim10 \times L_{gr}$, we can estimate that for a single layer of graphene $|\chi^{(3)}_{gr}|\sim10^{-7}$ esu.

It is interesting to compare the nonlinear optical properties of graphene to those of other carbon materials, such as carbon nanotubes \cite{12,13,14,15,16,27,28}. Carbon nanotubes can be purely metallic or possess a finite band gap, depending on the manner in which the edges of the graphene layer join to form the tube. Since nonlinear properties are known to be very dependent on the nature of electronic transitions \cite{16}, carbon nanotubes exhibit a broad range of nonlinear polarizabilities in the optical and infrared range. For thin films of nanotubes, $|\chi^{(3)}|\sim10^{-6}-10^{-8}$ esu has been reported \cite{12,13,14,15}. In particular, four-wave mixing measurements, similar to those carried out here, gave $|\chi^{(3)}|\sim10^{-8}$ esu \cite{15} for nanotube films. Due to the singularity in electron density of states at the band edges of nanotubes (Van Hove singularity) \cite{16,27,28}, theoretical predictions suggest that homogenous ensembles of identical tubes will exhibit $|\chi^{(3)}|$ which varies over six orders of magnitude on varying excitation wavelengths. This strongly resonant nature has a typical resonance width $\Delta \lambda < 1$ nm \cite{16}.

While our estimate of the nonlinear susceptibility of graphene is comparable in magnitude to that measured for nanotubes, we have shown that it is relatively constant with wavelength, Figure \ref{fig2}. We suggest that this frequency independence in the nonlinear response could be useful in graphene applications where a constant response is important, for example, in such nonlinear devices as bistable mirrors for mode-locking, optical limiters and switchable elements.

In summary, we have demonstrated that graphene exhibits a wavelength independent, strongly nonlinear optical response in the near infrared spectral region, described by a third-order susceptibility $|\chi^{(3)}|\sim10^{-7}$ esu, and shown that nonlinear measurements are a useful tool for visualizing graphene flakes on dielectric substrates. Studies of nonlinear properties of graphene represent a promising direction for optical experiments on this new material. It would be interesting to extend these studies to a broad range of frequencies, including the low-energy (THz) range. Experimental studies of nonlinear optical properties of graphene should be carried out in conjunction with theoretical investigations, which are currently very limited \cite{10}.

This work was funded by the RCUK and EPSRC, S.A.M. acknowledges financial support from the Deutsche Forschungsgemeinschaft. The authors also wish to thank W.L. Barnes, M. Bonn, M.J. Lockyear and A.F. Wyatt for helpful discussions.\\
\\
$^{\ast}$Electronic mail: E.Hendry@ex.ac.uk
 \newpage

\end{document}